\documentclass[prd,twocolumn,showkeys,floatfix,footinbib]{revtex4}
\usepackage{graphicx} 
\usepackage{amssymb} 
\usepackage{amsmath}
\usepackage{epsfig}
\newcommand{\slashed}[1]{#1\hspace*{-4.3pt}\slash}

\newcommand \Pomeron {I\!\!P}
\newcommand{\trans}{\ensuremath{t}}
\newcommand{\prob}{\ensuremath{\chi}}

\begin{document}
\title{Multiple Hard Partonic Collisions with Correlations in Proton-Proton Scattering} 
\author{T.~C.~Rogers}
\affiliation{Department of Physics and Astronomy,\\ Vrije Universiteit Amsterdam,\\ NL-1081 HV Amsterdam, The Netherlands}
\author{M.~Strikman}
\affiliation{Department of Physics,\\ Pennsylvania State University,\\ University Park, PA  16802, USA }

\begin{abstract}
We propose a simple method for incorporating  
correlations into the impact parameter space description of multiple (semi-)hard 
partonic collisions in high energy hadron-hadron scattering. 
The perturbative QCD input is the 
standard factorization theorem for inclusive dijet production with a lower cutoff on transverse momentum.
The width of the transverse distribution of hard 
partons is fixed by parameterizations of the two-gluon form factor.  
We then reconstruct the hard contribution to the total 
inelastic profile function and obtain corrections due to correlations to the more commonly used eikonal description.
Estimates of the size of double correlation corrections are based on the 
rate of double collisions measured at the Tevatron.
We find that, if typical values for the lower transverse momentum cutoff are used in the calculation of the 
inclusive hard dijet cross section, then the correlation corrections are necessary for maintaining 
consistency with expectations for the total inelastic proton-proton cross section at LHC energies. 
\end{abstract}

\keywords{perturbative QCD, unitarization, multiple collisions}
\maketitle

\section{Introduction}
\label{intro}

\begin{figure*}
\centering
\includegraphics[scale=.45]{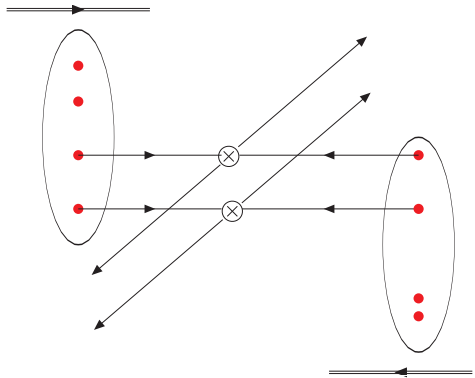}
\caption{Schematic depiction of a double hard collision.  A disconnected pair of partons from each proton collide to 
produce a pair of high transverse momentum dijets.}
\label{fig:doublescat}
\end{figure*}

Models of multiple partonic collisions in high energy hadron-hadron scattering are important for 
simulations of complex events in upcoming experiments like those at the LHC or in high energy cosmic ray air showers.
Furthermore, measurements of the rate of multiple partonic collisions can be used to test
current models of proton structure.
Already, measurements at accessible 
energies~\cite{Abe:1997xk,Abe:1993rv,Alitti:1991rd,Akesson:1986iv,Sonnenschein:2009nh}
yield a much smaller effective cross section $\sigma_{\rm eff}$ than what is naively expected if 
partons are homogeneously distributed over the transverse area of the proton.
(Figure~\ref{fig:doublescat} shows a schematic depiction of a double partonic collision.)
The definition of $\sigma_{\rm eff}$ is 
\begin{equation}
\label{eq:eff}
\sigma_{\rm eff} = m \frac{\sigma_2^2}{2 \sigma_4}
\end{equation}  
where $\sigma_2$ is the inclusive cross section for a single partonic collision (resulting in a dijet), 
$\sigma_4$ is the inclusive cross section for a double collision, and $m$ is a symmetry factor that depends on whether the 
partons are identical.
 New measurements of multiple collisions are currently being proposed for the LHC~\cite{Maina:2009vx}.
Hence, novel new phenomena involving multiple hard partonic collisions, which will enhance understanding of 
proton structure, can be expected in the next 
generation of experiments at the high energy frontier. 

However, multiple interactions involve a complex interplay of soft, hard, 
and semihard physics, so a complete description using purely 
perturbative techniques is not possible.  In simulations of complex hadronic final states, 
methods are needed for combining hard and soft collisions in a consistent way. 
(For an overview of current approaches see, e.g.~\cite{Engel:2003ac,Sjostrand:2004pf}, and references therein.)   
The hard contribution is calculated using 
the well-known perturbative QCD (pQCD) leading twist factorization formula for the
inclusive dijet production cross section, involving a convolution of the standard parton distribution 
functions (PDFs) with a partonic cross section.  
An immediate complication is that, in order for perturbation theory to be applicable, 
the relative transverse momentum of the produced jets 
must be larger than some minimum cutoff scale $p_t^c$. 
The cutoff should be chosen small so as to maximize the range of the perturbative expression,
but still large enough for perturbative methods to be reasonable.  
For describing events with transverse momentum less than $p_t^c$, nonperturbative methods are needed.
A prescription for matching the high and low transverse momentum behavior is necessary for a complete description over the full range of 
transverse momentum.

The inclusive pQCD hard dijet cross section is numerically very sensitive to the precise choice of $p_t^c$.  This has a tendency 
to lead to substantial variations between different model predictions of the minijet cross section at high energies. 
The question of 
what values of $p_t^c$ are appropriate continues to be discussed in current research on the 
development of models and simulations (see, for example, recent discussions 
in~\cite{Sjostrand:2004pf,Bahr:2008dy,Ahn:2009wx,Jung:2009eq}).
%%%%%%%%%%%%%%%%%%%%%%%%%%%%%%%%%%%
%%%%%%%%%%%%%%%%%%%%%%%%%%%%%%%%%%%

An additional complication is that a description is needed for the distribution of hard 
partons in impact parameter space.  In the past, it has usually been modeled or assumed to be
equal to the electromagnetic form factor of the nucleon.
Also, in a number of models, soft and hard partonic interactions are incorporated into a single eikonal picture.
In this way, both soft and hard interactions are 
included in a way that respects $s$-channel unitarity.  Data for the total 
and inelastic $pp$ cross sections can then be used to fit parameters such as $p_t^c$ and the width of the distribution 
of hard partons in impact parameter space.  However, different 
choices for these parameters can provide equally good fits to the total cross section at accessible energies while leading
to very different extrapolations at high energies (see, for example, Ref.~\cite{Engel:2001mm}).
Therefore, it is important to make use of any experimental or observational 
information that can narrow the range of allowed parameters and falsify some of the models 
currently in use.
 
It is nowadays possible to use pQCD to obtain direct experimental 
information about the transverse spread of hard partons in the 
proton via parametrizations of the generalized parton distributions (GPDs).  
(See Ref.~\cite{Weiss:2009ar} and references therein for a review 
of the phenomenology of GPDs.)  
The gluon GPD of the proton, for example, can be extracted 
from measurements of the $t$ dependence in 
deep inelastic production of light vector mesons 
or photoproduction of heavy vector mesons~\cite{Frankfurt:2002ka,Aktas:2005xu}.  
Then,  since the GPD is a universal object~\cite{Collins:1996fb}, it can 
be reused in other processes.  
In particular, it can be used in the description of the impact parameter dependence 
of hard collisions in hadron-hadron scattering.  
As such, we adopt the point of view in this paper that the impact parameter description of a hard collision
is not an adjustable model parameter, but rather is fixed by fits to the gluon GPD.

An accurate description of the impact parameter dependence is important because it allows for a measure of how close 
the hard interaction is to the unitarity (or ``saturation'') limit.
Furthermore, the proximity to the unitarity limit is related to the number 
of high transverse momentum jets that are produced, and hence to the use of centrality as a dijet trigger~\cite{Frankfurt:2003td}. 
However, as was illustrated in Ref.~\cite{Rogers:2008ua}, using GPDs to describe the impact parameter dependence 
of hard collisions within the  
approximation where partons are not correlated in the transverse plane
(which is implemented in the simplest eikonal description) leads  
to inconsistencies with general expectations for the extrapolation of 
the total inelastic profile function to high energies, unless 
a large value is used for the transverse momentum cutoff $p_t^c$.    
Specifically, the contribution to the total inelastic profile function from the production of hard dijet pairs 
becomes larger than the total inelastic profile function itself.
Although the inclusive dijet cross section is unitarized in the basic eikonal description, 
it nevertheless grows too rapidly with energy. 
As we will argue, this problem is most likely a symptom of the common assumption that partons are uncorrelated.

The study of correlations in multiple hard collisions 
is already an active area of research~\cite{Calucci:1999yz,Calucci:2008jw,Snigirev:2008pn,Domdey:2009bg}.
Implementations of Dokshitzer-Gribov-Lipatov-Alterelli-Parisi (DGLAP) 
evolution in multiparton distribution functions suggest that correlations are indeed significant~\cite{Snigirev:2008pn}.
Correlations may also be induced by evolution in Bjorken $x$~\cite{Mueller:1992pm} in the very high energy limit. 
However, numerical estimates in Ref.~\cite{Rogers:2008ua} suggest that  
hard unitarity (saturation) effects contribute to only a small fraction of the total inclusive dijet cross section, even at LHC energies. 
Moreover, the inconsistencies encountered in Ref.~\cite{Rogers:2008ua} occur even at large impact parameters, $\gtrsim 1.0$~fm, 
and values of $p_t^c$ that are not particularly small, e.g. $p_t^c \approx 2.5$~GeV.
In the instanton liquid model (see Ref.~\cite{Diakonov:2002fq} and references therein) one can expect a 
strong correlation between the quark and gluon fields in the constituent
quarks which will result in correlations at all impact parameters. It may be somewhat 
diluted at large $b$ due to the contribution from the $4q\bar q$ component. On the other 
hand for partons at large enough distances from the center it is likely that 
another parton should be present at a distance comparable to the confinement 
scale, leading to correlations of  $q\bar q$ pairs at 
relative distances $\le  \mbox{0.5 fm}$ that are rather small
which could easily compensate for the dilution effect.
Therefore, it is likely that nonperturbative correlations, unrelated to the overall impact parameter, 
play a role in determining the size of the dijet contribution 
to the total inelastic cross section.
Heuristically, one can imagine scattering at large impact parameters as the scattering of pion clouds.  
The nonperturbative dynamics 
responsible for binding the $q\bar{q}$ pairs introduces potentially large correlations.

Collinear constituents of the incoming protons are expected, on average, to have a 
spacetime separation of order $\sim 1/\Lambda_{\rm QCD}$.  
Because of the breakdown of asymptotic freedom at these scales, 
it should be anticipated that the constituents of the nucleon are subject to strong nonperturbative 
correlations. 
In other words, one should not expect the multiparton distribution functions to be
simply related to products of the single parton distribution function at any scale.

As a specific illustration, let us consider the double parton event in Fig.~\ref{fig:doublescat}.  
If $f_N(x_1,x_2)$ is the distribution function for parton pairs with momentum fractions $x_1$ and $x_2$ inside proton $N$, 
and $f_N(x_1)$ and $f_N(x_2)$ are the standard PDFs, then an often made ansatz is 
\begin{equation}
\label{eq:badapprox}
f_N(x_1,x_2) \stackrel{??}{\approx} f_N(x_1) \, f_N(x_2).
\end{equation}
In some treatments of correlations  
this relation follows after an integration over impact parameters, even if the factorization 
ansatz is broken in impact parameter space.
The question marks in Eq.~(\ref{eq:badapprox}) indicate that 
we will question the validity of this assumption.
Given the strong binding between the constituent partons in the proton, the approximation in Eq.~(\ref{eq:badapprox}) is probably 
very rough at any scale.  We will argue that deviations can have an 
important effect on extrapolations to high energies. 
 
For a large cutoff $p_t^c$, the inclusive dijet cross section 
gives a small fraction of the total inelastic cross section, and there is no conflict with $s$-channel unitarity.
What is needed at smaller $p_t^c$ is a method for organizing correlation corrections which does 
not rely on the commonly used assumption that correlations can be neglected. 

To summarize, the aim of this paper is to 
set up a method for organizing corrections to the 
uncorrelated 
approach at small $p_t^c$.
We will use this to 
simultaneously incorporate the following information 
into a description of multiple hard partonic collision:  
%%%%%%%%%%%%%%%%%%%%%%%%%%%%%%%%%%%%%%%%%%%%
\begin{itemize}
\item The impact parameter distribution of hard partons 
obtained from measurements of the gluon GPD.
The width of the distribution of hard partons is not a fixable parameter in our approach.
\item An estimate of the size of double parton correlations
obtained from measurements $\sigma_{\rm eff}$.  
In this paper, we will assume that these correlations are roughly 
independent of impact parameter.
\end{itemize}
From this information, we will reconstruct the hard dijet contribution to the 
total inelastic profile function. 
Compared with the  
uncorrelated 
expression, the result obtained with correlations will be shown to exhibit greater consistency 
with common high energy extrapolations of the total inelastic 
cross section, even with a relatively small and fixed value for $p_t^c$ in the perturbative inclusive dijet calculation.  
Thus, including correlations in this way 
may provide
a natural resolution to the 
consistency problems encountered in Ref.~\cite{Rogers:2008ua} at large
impact parameters and fixed $p_t^c$.

The paper is organized as follows:
In Sec.~\ref{sec:imp} we review the basic setup for discussing high energy collisions
in impact parameter space, and in Sec.~\ref{sec:multhard} we review the steps outlined in 
Ref.~\cite{Rogers:2008ua} for dealing with multiple hard collisions in terms of inclusive dijet cross sections, applying the result 
to the special case of uncorrelated scattering in Sec.~\ref{sec:uncorr}.
In Sec.~\ref{sec:corr} these steps are extended to allow for impact parameter independent correlations.
In Sec.~\ref{sec:numresults} we use estimates of correlations based on Tevatron measurements of $\sigma_{\rm eff}$ to calculate the hard
contribution to the inelastic profile function at LHC energies, and we compare with standard extrapolations of the 
total inelastic profile function.  
We speculate on prospects for including impact parameter dependence of correlations in Sec.~\ref{sec:impdep}.
We close in Sec.~\ref{sec:con} with a conclusion and a discussion of the main results.

\section{Impact Parameter Representation}
\label{sec:imp}

\subsection{S-Channel Unitarity}
\label{sec:impa}

%%%%%%%%%%%%%%%%%%%%%%%%%%%%%%%%%%% 
\begin{figure}
\centering
\includegraphics[scale=.5]{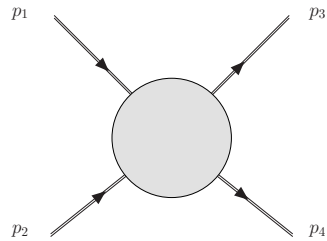}
\caption{Momentum labels for $pp$ scattering.}
\label{fig:kinematics}
\end{figure}
%%%%%%%%%%%%%%%%%%%%%%%%%%%%%%%%%%%
We work in impact parameter space, defining the profile function,
\begin{equation}
\label{profile}
\Gamma(s,b) \equiv \frac{1}{2 i s(2 \pi)^2} \int d^2{\bf q} \,e^{i {\bf q} \cdot {\bf b}} A(s,t), 
\end{equation}
Here $A(s,t)$ is the amplitude for elastic $pp$ scattering (see Fig.~\ref{fig:kinematics}), and 
$s$ and $t$ are the usual Mandelstam variables, $s = (p_1 + p_2)^2$ and $t = (p_1 - p_3)^2$.
We work in the high energy limit, $s >>  -t$ where one may approximate $t \approx -{\bf q}^2$
(see e.g. Ref.~\cite{BP02} for a review of kinematics in the high energy limit).
In the two-dimensional Fourier transform to coordinate space, ${\bf b}$ is the impact parameter. 

Unitarity and analyticity allow the total, elastic and inelastic cross sections to be calculated in 
terms of the profile function via the familiar relations:
\begin{eqnarray}
\label{eq:pp_profile}
\sigma_{tot}(s) & = & 2 \int d^2 {\bf b} \, {\rm Re} \, \Gamma(s,b), \label{eq:sigtot} \\
\sigma_{el}(s) & = & \int d^2 {\bf b} \, \left| \Gamma(s,b) \right|^2, \label{eq:sigel} \\
\sigma_{inel}(s) & = &\int d^2 {\bf b} \left( 2\,{\rm Re}  \,\Gamma(s,b) 
-  \left| \Gamma(s,b) \right|^2 \right). \; \label{eq:siginel} \,
\end{eqnarray} 
We refer to the integrand of Eq.~(\ref{eq:siginel}) as the inelastic profile function,
\begin{equation}
\label{eq:gammainel}
\Gamma^{\rm inel}(s,b) \equiv \left( 2\,{\rm Re}  \,\Gamma(s,b) 
-  \left| \Gamma(s,b) \right|^2 \right).
\end{equation}
In the very high energy limit, it is appropriate to neglect the imaginary  
part of the amplitude.  Then unitarity requires
\begin{equation}
\Gamma^{\rm inel}(s,b), \Gamma(s,b) \leq 1.
\end{equation}
The profile function for the total proton-proton cross section and its extrapolation to 
high energies has been studied extensively over the last few decades.  
Our main concern in this paper is whether common extrapolations of Eq.~(\ref{eq:gammainel}) are 
consistent with extrapolations of the total hard contribution.

\subsection{Inclusive Dijet Cross Section}
\label{sec:impb}

Hard collisions are described by the leading twist pQCD expression for the inclusive dijet cross section, 
\begin{multline}
\label{dijetformula}
\sigma_{\rm pQCD}^{\rm inc}(s;p_{t}^{c}) =  
\sum_{i,j,k,l} \frac{K}{1 + \delta_{kl}} \int d \, x_1 d \, x_2 \int d \, p_{t}^{2} \times \\ \times \,
\frac{d \hat{\sigma}_{i j \rightarrow k l}}{d p_{t}^{2}} \,
f_{i/p_1} (x_1 ; p_t)\, f_{j/p_{2}} (x_2 ; p_t)\, \theta(p_\trans - p_{t}^c)\; .
\end{multline}
The collision is between parton types $i$ and $j$ inside protons $p_1$ and $p_2$ respectively, and 
the partonic hard scattering cross section $d \hat{\sigma}_{i j \rightarrow k l} / d p_{t}^{2}$ is calculated at tree level. 
We have explicitly included a $K$ factor and a symmetry factor $1 / (1 + \delta_{kl})$.
The $f_{i/p_1} (x_1 ; p_t)$ and $f_{j/p_{2}} (x_2 ; p_t)$ are the usual integrated 
parton distribution functions for partons with 
longitudinal momentum fractions $x_1$ and $x_2$, evaluated at a hard scale equal to the dijet transverse momentum.
The lower bound $p_t^c$ is in principle arbitrary, but it should be chosen 
large enough for it to be a reasonable hard scale. 
It is not clear exactly what is the minimum $p_t^c$ that can be used, but Relativistic Heavy Ion Collider (RHIC) data for pion production 
suggest that perturbation theory is still reliable 
for pions with $p_t^c \gtrsim 1$~GeV~\cite{d'Enterria:2009am}
for forward production where background from soft physics is small.
(Note here that for such kinematics $p_t$ of the progenitor  quark is close to $p_t$ of  the pion.)

Information about the impact parameter distribution of hard partons in the proton is obtained from 
the gluon GPD which
is parametrized in experimental measurements of the $t$ dependence in exclusive heavy vector meson photoproduction 
and light vector meson electroproduction at small $x$.  
For the $t$ dependence of the differential exclusive vector meson production cross section, we use the dipole parametrization,
\begin{equation}
\label{eq:2glu}
F_g(x,t;\mu) = \frac{1}{\left( 1 - \frac{t}{m_g(x,\mu)^2} \right)^2}.
\end{equation}
The basic Feynman diagram contributing to this reaction involves the exchange of two gluons in the $t$ channel, so we refer to it 
as the \emph{two-gluon form factor}.  The gluon GPD $g(x,t;\mu)$ is then
\begin{equation}
\label{eq:gluGPD}
g(x,t;\mu) = g(x;\mu) F_g(x,t;\mu),
\end{equation}
where $g(x;\mu)$ is the standard integrated gluon distribution function evaluated at hard scale $\mu$.
The two-gluon form factor obeys the condition,
\begin{equation}
F_g(x,t=0;\mu) = 1
\end{equation}
so Eq.~(\ref{eq:gluGPD}) reduces to the standard PDF in the limit of $t \to 0$.
The parameter $m_g(x,\mu)$ in Eq.~(\ref{eq:2glu}) determines  
the width of the peak around $t = 0$.  Following 
Ref.~\cite{Frankfurt:2003td}, we allow
it to have $x$ and $\mu$ dependence to account for evolution in the hard scale $\mu$ and diffusion at small $x$.
The Fourier transform of the two-gluon form factor into the transverse plain is
\begin{equation}
\mathcal{F}_g(x,b;\mu) = \int d^2 {\bf \Delta_t} \, F_g(x,t;\mu) e^{-i {\bf \Delta_t} \cdot {\bf b} }, \qquad t \approx -{\bf \Delta_t^2}. 
\end{equation}
Again, the approximation $t \approx -{\bf \Delta}_t$ is justified so long as $s >> -t$.
Using the dipole form in Eq.~(\ref{eq:2glu}) one finds explicitly
\begin{equation}
\label{eq:2glub}
\mathcal{F}_g(x,b;\mu) = \frac{m_g(x;\mu)^3 b}{4 \pi} K_{1}(m_g(x;\mu) b).
\end{equation}
In this paper, $K_{n}$ for integer $n$ denotes a modified Bessel function of the second kind.
The overlap function 
is defined as
\begin{multline}
P_2(b,x_1,x_2;\mu) = \\ \int d^2 {\bf b^\prime} \, \mathcal{F}_g(x_1,| {\bf b^\prime} |;\mu) \, \mathcal{F}_g(x_2,| {\bf b - b^\prime} |;\mu).
\end{multline}
Using Eq.~(\ref{eq:2glub}) yields
\begin{equation}
\label{eq:FSWoverlap}
P_2(s,b;p_t^c) = \frac{m_g^2(x;p_t^c)}{12 \pi} \left( \frac{m_g(x;p_t^c) b}{2} \right)^3 K_3(m_g(x;p_t^c) b).
\end{equation} 
Here we have made the usual approximation, $x_1 \approx x_2 \approx x \equiv 2 p_t^c/ \sqrt{s}$.
A more detailed treatment should take into account the separate integrations over $x_1$ and $x_2$ ---
there is not, in general, a one-to-one mapping between values of $p_t^c$ and $x_1(2)$.
For now, we mention that direct numerical calculations verify that this approximation 
introduces less than $10\%$ error in the essential region of integration for the cross section.
Note that $P_2(s,b;p_t^c)$ is normalized to unity,
\begin{equation}
\label{eq:norm}
\int d^2 {\bf b} \; P_2(s,b;p_t^c) = 1.
\end{equation}
Combining Eqs.~(\ref{eq:2glu}-\ref{eq:norm}) with Eq.~(\ref{dijetformula}) allows the 
inclusive dijet cross section to be written in the form
\begin{equation}
\label{eq:bform}
\sigma^{\rm inc}_{\rm pQCD}(s;p_t^c) = \int d^2 {\bf b} \, \prob_2(s,b;p_t^c)
\end{equation}
where
\begin{equation}
\label{eq:prob_one}
\prob_2(s,b;p_t^c) = \sigma^{\rm inc}_{\rm pQCD}(s;p_t^c) P_2(s,b;p_t^c).
\end{equation}
We will refer to $\prob_2(s,b;p_t^c)$ as the impact parameter dependent inclusive cross section~\footnote{$\prob_2(s,b;p_t^c)$ was called 
$\mathcal{N}_2$ in Ref.~\cite{Rogers:2008ua}.  
We use $\prob_2$ here to be more consistent with the standard notation.  We avoid using $\Gamma$ in order 
to prevent confusion with the total profile function for $pp$ scattering.}.
Using the GPD to write the inclusive dijet cross section in the form of Eq.~(\ref{eq:bform}) enables one 
to analyze the contribution from different regions of impact parameter space.

\section{Multiple Hard Partonic Collisions}
\label{sec:multhard}

By taking into account multiple hard scattering events, it is in principle possible to reconstruct 
the total hard scattering contribution to the total inelastic profile function using 
probabilistic arguments~\cite{Ametller:1987ru,Capella:1981ju}.
In this section, we briefly review the steps 
for constructing the dijet contribution to the total inelastic profile function 
in terms of a series involving the inclusive $n$-dijet cross sections.

We start by defining $\prob_{2 n}(s,b;p_t^c)$ to be the analogue of $\prob_{2}(s,b;p_t^c)$ 
for the case of $n$-dijet production ($n$ hard collisions).
Namely, integrating over all impact parameters yields
\begin{equation}
\int d^2 {\bf b} \; \prob_{2n}(s,b;p_t^c) = \sigma^{inc}_{2n}(s;p_t^c),
\end{equation}
where $\sigma^{inc}_{2n}(s;p_t^c)$ is the inclusive cross section for producing $n$-dijet pairs.

Next, $\tilde{\prob}_{2n}(s,b;p_t^c)$ is defined to be the exclusive analogue of $\prob_{2 n}(s,b;p_t^c)$.  
It describes the production of \emph{exactly} $n$ hard collisions, differential in $b$.  Integrating over impact parameters gives
\begin{equation}
\int d^2 {\bf b} \; \tilde{\prob}_{2n}(s,b;p_t^c) = \sigma^{ex}_{2n}(s;p_t^c),
\end{equation}
where $\sigma^{ex}_{2n}(s;p_t^c)$ is the integrated cross section for producing \emph{exactly} $n$-dijet pairs.
[The $\tilde{\prob}_{2n}(s,b;p_t^c)$ correspond to the ``exclusive'' cross sections of Ref.~\cite{Calucci:2008jw}, the quotes 
referring to the fact that these cross sections are still inclusive in soft fragments.]

Now we reconstruct the hard dijet contribution to the inelastic profile function 
by writing down the expression for the inclusive cross section for $k$-dijet production in terms of the exclusive cross sections:
\begin{equation}
\label{eq:incl}
\prob_{2k}(s,b;p_t^c) = \sum_{n \geq k}^{\infty} \, {n \choose k}  \tilde{\prob}_{2n}(s,b;p_t^c).
\end{equation}
The combinatorial factor counts all the ways an $n$-dijet event can contribute to the inclusive $k$-dijet cross section.
Equation~(\ref{eq:incl}) can be inverted to obtain the exclusive impact parameter cross sections in terms of the inclusive ones,
\begin{equation}
\label{eq:ex}
\tilde{\prob}_{2k}(s,b;p_t^c) = \sum_{n \geq k}^{\infty} \, {n \choose k} (-1)^{n-k} \, \prob_{2n}(s,b;p_t^c).
\end{equation}
The total inelastic profile function is obtained by summing all the exclusive components.  Using Eq.~(\ref{eq:ex}), we obtain
\begin{multline}
\label{eq:inel_profile}
\Gamma_{\rm dijets}^{\rm inel}(s,b;p_t^c) = \sum_{k=1}^{\infty} \tilde{\prob}_{2k}(s,b;p_t^c) = \\ =
\sum_{n=1}^{\infty} (-1)^{n-1} \prob_{2n}(s,b;p_t^c).
\end{multline}
In principle, if all the inclusive $n$-dijet impact parameter dependent cross sections $\chi_{2n}(s,b;p_t^c)$ are known, 
then it is possible to obtain exactly the dijet contribution  
to the inelastic profile function by summing all the terms 
in the last line of Eq.~(\ref{eq:inel_profile}).  
In practice, higher orders in $n$ 
need to be modeled or approximated.  

Consistency requires the hard dijet contribution to the total inelastic cross section to be less than the 
actual total inelastic cross section so 
\begin{equation}
\label{eq:consistency}
\Gamma^{\rm inel}_{\rm dijets}(s,b;p_t^c) \leq \Gamma_{\rm actual}^{\rm inel}(s,b),
\end{equation}
where the right side is the ``actual'' inelastic profile function, which could be obtained from either a measurement or a model extrapolation.
Hence, Eq.~(\ref{eq:consistency}) provides a means of checking that an expression for $\Gamma^{\rm inel}_{\rm dijets}(s,b;p_t^c)$, constructed from 
Eq.~(\ref{eq:inel_profile}), is consistent with other methods for obtaining the total inelastic profile function.  A violation of 
Eq.~(\ref{eq:consistency}) means either that the model/extrapolation is incorrect or that there is a problem with the $\prob_{2n}(s,b;p_t^c)$ 
used in Eq.~(\ref{eq:inel_profile}).
 
\section{Uncorrelated Scattering}
\label{sec:uncorr}

\begin{figure*}
\centering
\includegraphics[scale=.9]{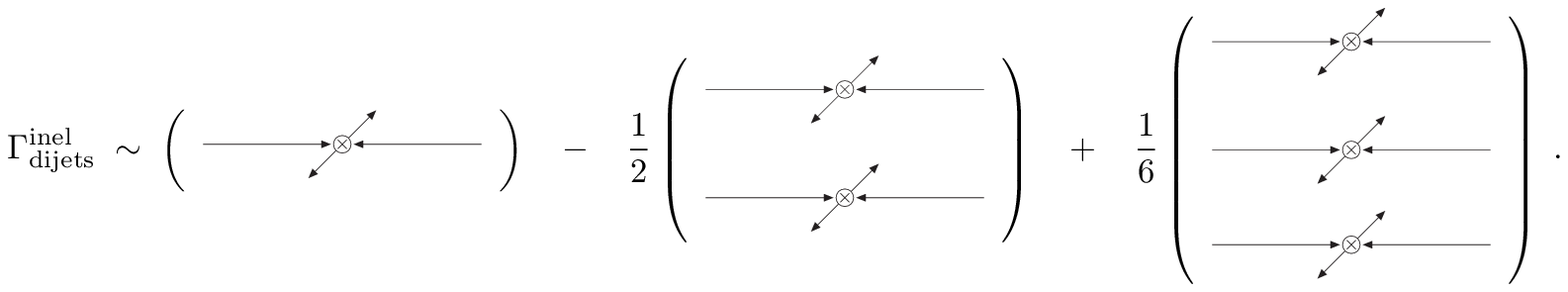}
\caption{Graphical representation of the term in the series for uncorrelated scattering --- the first three terms in the second line of Eq.~(\ref{eq:profile}), assuming no correlations.  Spectator partons are not shown.}
\label{fig:uncorrelated}
\end{figure*}
%%%%%%%%%%%%%%%%%%%%%%%%%%%%%%%%%%%
The simplest and most common way to obtain an explicit expression for $\Gamma^{\rm inel}_{\rm dijets}(s,b;p_t^c)$ from Eq.~(\ref{eq:inel_profile}) is to 
assume that all partonic collisions occur completely independently from one another.  
It was shown in Ref.~\cite{Ametller:1987ru} that the inclusive impact parameter dependent cross section for production of $n$ dijets is then
\begin{equation}
\label{eq:naive_prob}
\prob_{2 n}(s,b;p_t^c) = \frac{1}{n!} \prob_2(s,b;p_t^c)^n.
\end{equation}
This can be inserted into the last line of Eq.~(\ref{eq:inel_profile}) and summed to reproduce
the familiar unitarized eikonal-like expression,
\begin{multline}
\label{eq:profile}
\Gamma_{\rm dijets}^{\rm inel}(s,b;p_t^c) = 
\sum_{n=1}^{\infty} \frac{1}{n!} (-1)^{n-1} \prob_2(s,b;p_t^c)^n = \\ =
1 - \exp \left[ - \prob_2(s,b;p_t^c) \right] \; .
\end{multline}
The single, double, and triple scattering terms are represented graphically in Fig.~\ref{fig:uncorrelated}. 
(This kind of graphical representation will be useful later for describing combinatorial factors when 
correlations are included.) 
Each circle-cross represents a hard scattering event.
The uncorrelated assumption of Eq.~(\ref{eq:naive_prob}) is symbolized by the absence of any lines
connecting the different hard collisions - each graph is simply $\chi_2(s,b)$ raised to the appropriate power.

For many practical purposes, Eq.~(\ref{eq:profile}) is sufficient.
In general, the reconstructed profile function simply needs to reproduce the correct pQCD expression at large $b$ where 
multiple collisions are very rare, while the minimal unitarity requirement that $\Gamma^{\rm inel}_{\rm dijets} \lesssim 1$ should
be enforced at small $b$.  That is, the basic requirements in the high energy limit are,
\begin{eqnarray}
\Gamma_{\rm dijets}^{\rm inel}(s,b) & \stackrel{b \to \infty}{=} &  \prob_2(s,b;p_t^c) \label{eq:basicreqs1} \\
\Gamma_{\rm dijets}^{\rm inel}(s,b) & \stackrel{b \to 0}{\lesssim} & 1. \label{eq:basicreqs2}
\end{eqnarray}
In the high energy limit, $\Gamma_{\rm dijets}^{\rm inel}(s,b)$ is expected to approach one at small $b$ (black disk limit).
Equation~(\ref{eq:profile}) is completely satisfactory as far as conditions~(\ref{eq:basicreqs1},\ref{eq:basicreqs2}) are concerned. 
For large $b$, only the first term in the series in Eq.~(\ref{eq:profile}) --- i.e. single scattering --- is important. 
However, in a precise treatment, one should also account for potential for violations of the consistency requirement in Eq.~(\ref{eq:consistency}) in 
the high energy limit and at intermediate values of $b$ where corrections of order $\prob_2(s,b;p_t^c)^2$ are non-negligible.  
Indeed, such consistency problems were found in Ref.~\cite{Rogers:2008ua}.

\section{Organizing Correlations in Multiple Collisions}
\label{sec:corr}

\subsection{Impact Parameter Independence}
\label{sec:impindep}

Deviations from uncorrelated scattering can arise from multiple sources.
As discussed in the Sec.~\ref{intro}, correlations can be generated both in perturbative 
evolution equations and in nonperturbative models.

Correlations will also be induced by kinematical constraints.  We will assume, however, that most 
active partons have small enough $x$ that these constraints are unimportant, at least for 
the first few terms in the series in Eq.~(\ref{eq:profile}).
For this paper, we will assume that the incoming partons that take part in multiple hard collisions 
move nearly parallel with transverse momentum of order $\sim 1 / \Lambda_{\rm QCD}$.
That is, they have momentum typical for bound constituents of the incoming hadrons.   
In general, if $p_t^c$ is allowed to be larger than a few GeV, the partons will undergo 
DGLAP evolution, and hence may include partons with larger transverse momentum. 
Furthermore, one expects significant dependence of $\sigma_{eff}$ on the hard scale 
at large $p_t^c$~\cite{Domdey:2009bg}.  In such cases, it is possible that correlations
may be understood as arising from parton evolution.
However, conflicts with Eq.~(\ref{eq:consistency}) become less likely at larger $p_t^c$.

Therefore, we organize the description of correlations around the assumption that the effect is 
to introduce a simple (impact parameter independent) rescaling from the 
uncorrelated case. 
As a first example, we reconsider double hard collisions. 
Equation~(\ref{eq:naive_prob}) gives the uncorrelated expression
\begin{equation}
\label{eq:doubuncorr}
\prob_{4}(s,b;p_t^c) = \frac{1}{2} \prob_2(s,b;p_t^c)^2
\end{equation}
which should be replaced in the correlated case by
\begin{equation}
\label{eq:doublecorr}
\prob_4(s,b;p_t^c) \to \frac{1}{2} \left( 1 + \eta_4(s) \right) \chi_2(s,b;p_t^c)^2,
\end{equation}
where $\eta_4(s)$ parametrizes the deviation from uncorrelated scattering.
Our strategy is to estimate the size of the double correlation by directly 
fitting Eq.~(\ref{eq:doublecorr}) to experimental data, given the constraint that $\chi_2(s,b;p_t^c)$ is fixed by 
the GPD in Eq.~(\ref{eq:prob_one}).   
Note that we place no condition on the $b$ integral of Eq.~(\ref{eq:doubuncorr}). 
In particular, we do not use the approximation in Eq.~(\ref{eq:badapprox}). 
In general, the correlation correction will also depend on both $p_t^c$ and $b$.
For our analysis, we will not explicitly write the $p_t^c$ arguments in Eq.~(\ref{eq:doubuncorr}) because 
we are mainly concerned with correlation corrections in the limited range of $p_t^c$ where Eq.~(\ref{eq:consistency})
becomes problematic within the usual eikonal picture.  As we will see, neglecting the $b$ dependence in $\eta_4(s)$ will 
allow for a direct parametrization of the correlation correction in terms of experimentally observed double scattering rates.  
It is likely that this is a very rough approximation, but it will allow for a first estimate of the role of correlations
at large impact parameters.  We also remark that the dynamics responsible for confinement are likely to induce large
correlations regardless of impact parameter.  We will discuss possible $b$ dependence in greater detail in Sec.~\ref{sec:impdep}.

In experiments the effect of double partonic collisions is most commonly represented by the observable,
\begin{equation}
\label{eq:sigeffdef}
\sigma_{\rm eff} = \frac{1}{2} \frac{\sigma_{2}^{inc}(s;p_t^c)^2}{\sigma_{4}^{inc}(s;p_t^c)}.
\end{equation}
In the uncorrelated case, using Eq.~(\ref{eq:doubuncorr}) and Eq.~(\ref{eq:prob_one}) in Eq.~(\ref{eq:sigeffdef}) yields 
\begin{equation}
\label{eq:sigeffuncor}
\sigma_{\rm eff}^{\rm uncor} = \frac{1}{\int d^2 {\bf b} \, P_2(s,b;p_t^c)^2 }.
\end{equation}
In general, the value of $\sigma_{\rm eff}$ can be fitted to experimentally measured values by changing the width or shape of $P_2(s,b;p_t^c)$.  
However, in our approach $P_2(s,b;p_t^c)$ is fixed by experimental measurements of the GPD, so the width of 
$P_2(s,b;p_t^c)$ is not a free parameter.  

If Eq.~(\ref{eq:doublecorr}) is used in Eq.~(\ref{eq:sigeffdef}), one obtains for $\sigma_{\rm eff}$ in the correlated case
\begin{equation}
\label{eq:sigeffcor}
\sigma_{\rm eff}^{\rm cor} = \frac{1}{(1 + \eta_4(s)) \int d^2 {\bf b} \, P_2(s,b;p_t^c)^2 }.
\end{equation}
With $P_{2}(s,b;p_t^c)$ fixed by the two-gluon form factor, one can only tune Eq.~(\ref{eq:sigeffcor}) 
to the measured value of $\sigma_{\rm eff}$ by adjusting $\eta_4(s)$.
Note also that, although we have assumed impact parameter independent correlations, the relative \emph{rate} of double collisions  
compared with single collisions, $\prob_2(s,b)/ \prob_4(s,b)$, still depends on impact parameter. 

\subsection{Double Partonic Correlations in Multiple Collisions}
\label{sec:doublemult}

%%%%%%%%%%%%%%%%%%%%%%%%%%%%%%%%%%% 
\begin{figure}
\centering
\includegraphics[scale=.9]{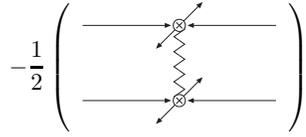}
\caption{Graphical representation of the extra contribution $(-1 /2) \eta_4(s) \chi_2(s,b;p_t^c)^2$ due to 
double partonic correlations in Eq.~(\ref{eq:doublecorr}).}
\label{fig:doublezigzag}
\end{figure}
%%%%%%%%%%%%%%%%%%%%%%%%%%%%%%%%%%%
%%%%%%%%%%%%%%%%%%%%%%%%%%%%%%%%%%%
\begin{figure*}
\centering
\includegraphics[scale=.9]{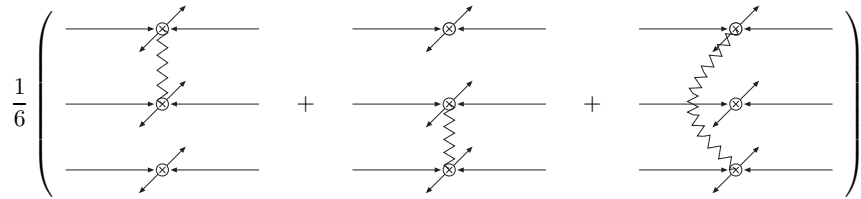}
\caption{Graphical representation of the extra contribution to the $n=3$ term of Eq.~(\ref{eq:profile}) due to double partonic correlations.}
\label{fig:triplezigzag}
\end{figure*}
%%%%%%%%%%%%%%%%%%%%%%%%%%%%%%%%%%%
The effect of double correlations in an $n$-collision event may now be organized in a very convenient way.  
We start by looking at how $\Gamma^{\rm inel}_{\rm dijets}(s,b;p_t^c)$ 
is modified by the inclusion of double correlations.
In Eq.~(\ref{eq:doublecorr}), $\eta_4(s)$ parametrizes the deviation of $\chi_4(s,b)$ from the uncorrelated case, $\eta_4(s) = 0$.
It represents a correction to the assumption in Eq.~(\ref{eq:badapprox}) that the integrated double parton PDF is simply a product 
of the standard PDFs.
The additional term proportional to $\eta_4(s)$ is represented by Fig.~\ref{fig:doublezigzag}.
The zigzag line connecting the two hard collisions may be thought of loosely as representing the effect of summing all soft gluons exchanged 
between the nearly parallel incoming and outgoing partons. 
We call $\eta_4(s)$ the double correlation correction factor.

Next, we reconsider the uncorrelated description of triple parton scattering, graphically represented by 
the third term in Fig.~\ref{fig:uncorrelated}.  
With no zigzag lines, we get the naive uncorrelated contribution from Eq.~(\ref{eq:naive_prob})
\begin{equation}
\chi_6(s,b;p_t^c) = \frac{1}{6} \prob_2(s,b;p_t^c)^3.  
\end{equation}    
For each pair of incoming partons there is another double 
correlation correction.
In other words, for each pair of colliding partons there is another replacement like Eq.~(\ref{eq:doublecorr}):
\begin{multline}
\chi_6(s,b;p_t^c) = \frac{1}{6} \; \left[ \prob_2(s,b;p_t^c)^2  \right] \; \prob_2(s,b;p_t^c) \\
\to \frac{1}{6} \; \left[ \left( 1 + \eta_4(s) \right) \chi_2(s,b;p_t^c)^2 \right] \; \prob_2(s,b;p_t^c).  
\end{multline} 
There is, therefore, an extra contribution equal to $\frac{\eta_4(s)}{6} (\prob_2(s,b))^3$ for 
each of the ${3 \choose 2} = 3$ ways a pair of 
incoming bound partons can become correlated.
This is illustrated graphically in Fig.~\ref{fig:triplezigzag}, which shows the additional contributions that must be 
added to the $n=3$ term in Eq.~(\ref{eq:profile})/Fig.~\ref{fig:uncorrelated}. 
The expression for $\chi_6(s,b;p_t^c)$ is therefore
\begin{equation}
\label{eq:tripsub}
\chi_6(s,b;p_t^c) = \frac{1}{6} \left( 1 + 3   \eta_4(s) \right) \prob_2(s,b;p_t^c)^3.  
\end{equation}

Following this example, 
it is now clear how to include double correlation corrections in $n$-parton scattering.  
In an $n$-parton collision, there are $n \choose 2$ 
additional contributions equal to $\frac{1}{n!} \eta_4(s) \chi_2(s,b;p_t^c)^n$.
In terms of diagrams like Fig.~\ref{fig:triplezigzag}, this corresponds to all the ways that 
two collisions can be connected by a single zigzag line.
Therefore, to include double correlations in the description of the inclusive $n$-dijet 
cross section, the uncorrelated relation in Eq.~(\ref{eq:naive_prob}) should be replaced with, 
\begin{multline}
\label{eq:new_prob}
\prob_{2n}(s,b;p_t^c) 
= \frac{1}{n!} \left( 1 + \eta_4(s) \frac{n(n-1)}{2} \right) \, \prob_2(s,b;p_t^c)^n.
\end{multline}
Using Eq.~(\ref{eq:new_prob}) in Eq.~(\ref{eq:inel_profile}) and summing over all $n$ 
produces an analytic expression for the hard dijet contribution to the total inelastic profile function,
\begin{multline}
\label{eq:profile2}
\Gamma_{\rm jets}^{\rm inel}(s,b;p_t^c) = \\
\sum_{n=1}^{\infty} \frac{(-1)^{n-1}}{n!} \left( 1 + \eta_4(s) \frac{n(n-1)}{2} \right) \, \prob_2(s,b;p_t^c)^n \\
= 1 - \exp \left[ - \prob_2(s,b;p_t^c) \right] - \\ - \frac{\eta_4(s)}{2} \, \prob_2(s,b;p_t^c)^2 \, \exp \left[ - \prob_2(s,b;p_t^c) \right]\; .
\end{multline}
Note that this equation respects the basic requirements of Eqs.~(\ref{eq:basicreqs1},\ref{eq:basicreqs2}) as long as $\eta_4(s) > 0$.
If $\eta_4$ is allowed to be less than zero, then there is a potential for 
Eq.~(\ref{eq:profile2}) to be greater than one for some intermediate impact parameters.
%%%%%%%%%%%%%%%%%%%%%%%%%%%%%%% 
The third line is the 
standard unitarized expression Eq.~(\ref{eq:profile}) for the profile function, familiar from the eikonal model, while the last 
line is a correction due to double correlations.  
The double correlation correction term 
contributes a power of $\chi_2(s,b)^2$ in a series expansion in small $\prob_2(s,b;p_t^c)$.  
Therefore, it will only become important at impact parameters which are small enough that terms proportional to $\chi_2(s,b;p_t^c)^2$ are non-negligible.

\subsection{Higher Correlations}

In the last section, we assumed that only a single pair of partons can become correlated.
This is reasonable if the goal is simply to account for double correlations at large or intermediate impact 
parameters where the contribution from double collisions [order $\chi_2(s,b)^2$] is significant, while contributions from 
triple collisions [order $\chi_2(s,b)^3$] and higher are negligible.
The explicit sum over all collisions in Eq.~(\ref{eq:profile2}) is needed to produce an analytic 
expression for the corrected inelastic profile function that still 
satisfies Eqs.~(\ref{eq:basicreqs1},\ref{eq:basicreqs2}) and is less than unity for all $b$.

\begin{figure*}
\centering
\includegraphics[scale=.3]{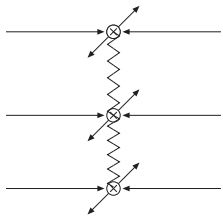}
\caption{Graphical illustration of triple correlations in triple parton scattering.}
\label{fig:triplecoll}
\end{figure*}

In reality, there are of course corrections from triple and higher correlations; 
in our graphical representation, triple correlations are represented by 
zigzag lines connecting three of the interaction points  --- see Fig.~\ref{fig:triplecoll}.  
%%%%%%%%%%%%%%%%%%%%%%%%%%%%%%%%%%%
The contribution from triple correlations becomes important only at order $\prob_2(s,b;p_t^c)^3$.
If powers of $\prob_2(s,b;p_t^c)^3$ are significant, then we can iterate the steps of 
Sec.~\ref{sec:doublemult} by replacing Eq.~(\ref{eq:tripsub}) with
\begin{equation}
\label{eq:tripsub2}
\chi_6(s,b;p_t^c) \to \frac{1}{6} \left( 1 + 3   \eta_4(s) + \eta_6(s) \right) \prob_2(s,b;p_t^c)^3, 
\end{equation}
exactly analogous to Eq.~(\ref{eq:doublecorr}) for double correlations.
The $\eta_6(s)$ parametrizes the correction from triple correlations in triple and higher partonic collisions.

Taking into account the contribution from triple correlations to $n > 3$ collisions, and including the 
appropriate combinatorial factors by counting all ways of connecting three hard collisions, we then 
recover Eq.~(\ref{eq:profile2}), but with a triple correlation correction term 
equal to $$\frac{ \eta_{6}(s)}{6} \, \prob_2(s,b;p_t^c)^3 \, \exp \left[ - \prob_2(s,b;p_t^c) \right].$$

Now it is a simple matter to generalize the steps from Sec.~\ref{sec:doublemult} to the 
arbitrary case of $n$ correlation corrections.  The resulting general expression for the inelastic profile function is  
\begin{multline}
\label{eq:general}
\Gamma_{\rm jets}^{\rm inel}(s,b;p_t^c) = \\
= 1 - \exp \left[ - \prob_2(s,b;p_t^c) \right] - \\ - \sum_{n=2}^{\infty} 
\frac{(-1)^n \eta_{2n}(s)}{n!} \, \prob_2(s,b;p_t^c)^n \, \exp \left[ - \prob_2(s,b;p_t^c) \right].
\end{multline}
The series after the first line includes all correlation corrections.  There is a new 
correlation correction factor $\eta_{2n}(s)$ for each number $n$ of collisions.
In the series representation for $\Gamma_{\rm jets}^{\rm inel}(s,b;p_t^c)$, 
a correlation correction factor $\eta_{2j}(s)$ is accompanied by powers $\prob_2(s,b;p_t^c)^j$ or higher. 

Equation~(\ref{eq:general}) always satisfies the basic conditions Eqs.~(\ref{eq:basicreqs1},\ref{eq:basicreqs2}) 
of a unitarized profile function.  [Although, depending on the signs of the $\eta_{2n}(s)$, it may need to be 
checked that the profile function does not exceed unity for some intermediate value of $b$.]
Contributions from higher $n$ correlation corrections are suppressed by factors of $(\prob_2(s,b))^n / n!$ and can be neglected so 
long as $\prob_2(s,b)$ is sufficiently smaller than 1.
By truncating the series at larger $n$, we obtain an increasingly refined description of the $b$ tail at moderate to large $b$.
By using models of multiple collisions to obtain the $\eta_{2n}(s)$, 
or directly parametrizing the size of correlation corrections from experimental data, it 
should therefore be possible to reconstruct an inelastic profile function that respects Eq.~(\ref{eq:consistency}).

Unfortunately, there is as yet very little direct experimental knowledge of $\eta_{2n}(s)$ for $n > 2$.  
However, in the next section we will argue that even when only double correlations are included,  
the corrections are important at moderate to large impact parameters. 
Once data are available, steps analogous to those in Sec.~\ref{sec:impindep} can be used to parametrize $\eta_6(s)$.
As in Ref.~\cite{Maina:2009vx}, we can define the triple effective cross section,
\begin{equation}
\label{eq:tripdef}
\left( \sigma_{\rm eff}^{T} \right)^2 = \frac{1}{6} \frac{\sigma_{2}^{inc}(s;p_t^c)^3}{\sigma_{6}^{inc}(s;p_t^c)}.
\end{equation}
Then, including up to triple correlations, we have
\begin{equation}
\label{eq:tripcalc}
\left( \sigma_{\rm eff}^{T} \right)^2 = \frac{1}{(1 + 3 \eta_4(s) + \eta_6(s)) \int d^2 {\bf b} \, P_2(s,b;p_t^c)^3 }.
\end{equation}
From Eq.~(\ref{eq:tripcalc}), we can calculate the correction from double correlations to $\sigma_{\rm eff}^{T}$ with 
triple correlations neglected.  In the next section, we will find values of $1.3$ or $2.1$ for $\eta_4(s)$ at currently accessible energies
and $p_t^c \approx 2.5$~GeV.  These give
$\sigma_{\rm eff}^{T} = 12.8$~mb and $10.5$~mb respectively, compared with $\sigma_{\rm eff}^{T} = 28.3$~mb for the case with no double correlations. 

We end this section by pointing out that, in principle, the steps leading to Eqs.~(\ref{eq:profile2},\ref{eq:general}) remain 
valid if we allow the $\eta_{2n}(s)$ to have impact parameter dependence.  The hypothesis of impact parameter independent correlations
is only needed if we wish to estimate the size of correlations by using Eq.~(\ref{eq:sigeffcor}).

\section{Numerical Estimates}
\label{sec:numresults}

\subsection{Total Cross Sections}
\label{sec:numtotcross}

Any numerical results that we obtain using Eq.~(\ref{eq:general})
should be compared with common extrapolations of the total inelastic profile function 
to high energies so that consistency with Eq.~(\ref{eq:consistency}) can be verified.  
A standard parametrization of the total profile function takes the form
\begin{equation}
\label{eq:totparam}
\Gamma(s,b) = \frac{\sigma_{tot}(s)}{4 \pi B(s)} \exp \left\{ -\frac{{\bf b}^2}{2 B(s)} \right\}
\end{equation}
with $B(s) \approx B_0 + \alpha^\prime \ln s$.  Regge theory fits give $\alpha \approx 0.25$~GeV$^{-2}$ for the rate of growth of $B(s)$.
For the LHC energy of $\sqrt{s} = 14$~TeV, a survey of common models and extrapolations in the 
literature~\cite{Block:1991yw,Khoze:2000wk,GPS,totem,Lami:2006ga} suggests the following 
as a range of reasonable parameters:
\begin{align}
90\, {\rm mb}  \, \, & \lesssim  & \sigma_{tot}(\sqrt{s} = 14\, {\rm TeV}) & \lesssim   & 120\, {\rm mb} \, \, \,   \label{eq:params1} \\
19\, {\rm GeV}^{-2} & \lesssim  & B(\sqrt{s} = 14\, {\rm TeV}) & \lesssim  & 23\, {\rm GeV}^{-2}.  \label{eq:params2}
\end{align}
For example, in Ref.~\cite{Frankfurt:2006jp} it is found that Eq.~(\ref{eq:totparam}) with $\Gamma(b = 0) = 1$ and 
$B = 21.8$~GeV$^{-2}$ is in very close agreement with the Regge parametrization of Ref.~\cite{Khoze:2000wk} as well as with the non-Gaussian 
model of Ref.~\cite{Islam:2002au}.
[Some fits put the maximum from the total upper error band for $\sigma_{tot}(\sqrt{s} = 14\, {\rm TeV})$ at around $130$~mb.  
However, a total cross section this large would also require a very large $B$ to avoid having a profile function greater 
than unity at small $b$.]

\subsection{Correlated vs uncorrelated partons}
\label{sec:numcorvsuncor}

To calculate $\sigma_{\rm eff}$ within the uncorrelated assumption, we
use Eq.~(\ref{eq:sigeffuncor}), and obtain $\prob_2(s,b;p_t^c)$ from the two-gluon form factor, as in Eq.~(\ref{eq:prob_one}).
We use $m_g \approx 1$~GeV which works well 
for $0.03 \leq x \leq 0.05$ and large $p_t$, relevant for most Tevatron data.    
More data on $J/\psi$ electro(photo)-production and deeply virtual Compton scattering in this range of kinematics would 
be very desirable for improving the accuracy of the determination of the $b$ dependence of quark and gluon GPDs
at $x \sim 10^{-2}$.

The value of $\sigma_{\rm eff}$ obtained from Eq.~(\ref{eq:sigeffuncor}) is then about $34$~mb. 
At small $x$, the width of the $\prob_2(s,b;p_t^c)$
grows and results in an even larger value for $\sigma_{\rm eff}$.  
The precise rate of growth of the radius at small $x$ and fixed $p_t$ is not currently 
well established but will likely become clearer as new data become available.
For the $x$ dependence, we will use the parametrization in Ref.~\cite{Frankfurt:2003td}.

The $34$~mb calculation obtained with the uncorrelated approximation 
should be compared with the measured value of $14.5$~mb~\cite{Abe:1997xk} from the CDF collaboration 
taken at a center-of-mass energy of $\sqrt{s} = 1.8$~GeV.
The uncorrelated calculation is roughly a factor of $2.3$ too large, implying
that it is unsafe to neglect corrections from correlations.
At a minimum one should keep the $\eta_4(s)$ term in Eq.~(\ref{eq:profile2}) with a correlation correction factor
$\eta_4 \approx 1.3$.  

It was argued in Ref.~\cite{Treleani:2007gi} that the analysis in~\cite{Abe:1997xk} actually overestimates $\sigma_{\rm eff}$.   
If three-jet events are taken into account (to make the cross section truly inclusive), then 
a new estimate is $\sigma_{\rm eff} \approx 11$~mb.  This suggests that the correlation correction is 
closer to $\eta_4 \approx 2.1$.  More recent measurements from the D$\slashed{0}$ collaboration 
find $\sigma_{\rm eff} = 15.1$~mb~\cite{Sonnenschein:2009nh}, without 
cuts on three-jet events.  So the precise size of $\sigma_{\rm eff}$ remains unclear.  We remark that $\sigma_{\rm eff}$ may depend on $p_t^c$, which
may lead to differences in measured values~\cite{Domdey:2009bg}.

\begin{figure*}
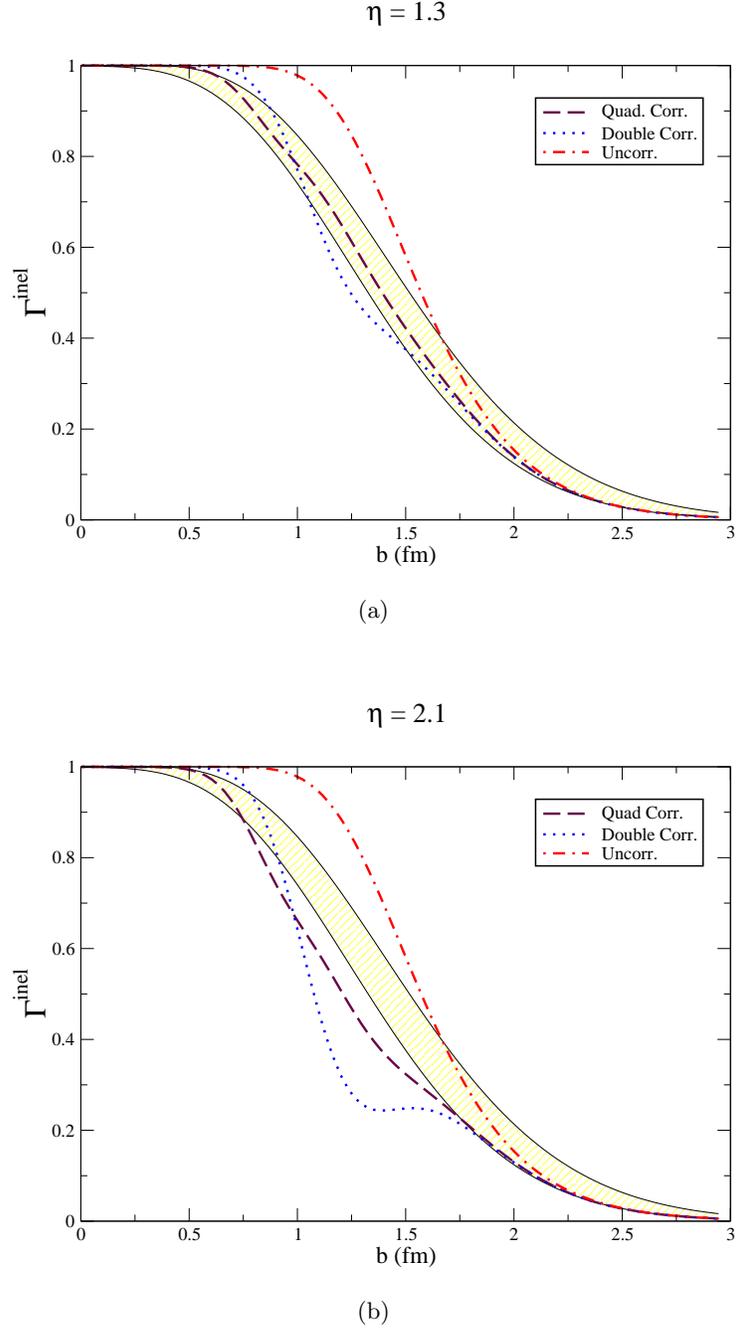

\centering
\begin{tabular}{c}
    \includegraphics[scale=0.4]{profeta1p3}
  \\[3mm]
  (a)
  \\[10mm]
    \includegraphics[scale=0.4]{profeta2p1}
  \\[3mm]
  (b)
  \end{tabular}
\caption{Inelastic profile functions calculated with and without correlations for $\sqrt{s} = 14$~TeV and 
$p_t^c = 2.5$~GeV.
The shaded region corresponds to the range of typical extrapolations.
The dash-dotted curve corresponds to the standard eikonal expression.
The dotted curve is the inelastic profile function including the double correlation correction 
in Eq.~(\ref{eq:profile2}) with (a) $\eta_4 = 1.3$~\cite{Abe:1997xk} and  (b) $\eta_4 = 2.1$~\cite{Treleani:2007gi}.
The dashed curve is with the triple quadruple and correlation corrections from Eq.~(\ref{eq:general}) 
using $\eta = \eta_4 = \eta_6 = \eta_8=$(a)$1.3$ and (b)$2.1$.}
\label{fig:corr_compare2}
\end{figure*}

The CDF measurements in~\cite{Abe:1997xk} find that correlations depend weakly on $x$, suggesting that $\eta_4$ may be roughly 
constant with energy.  Therefore, we will test the effect of using $1.3 \lesssim \eta_4(\sqrt{s} = 14\, {\rm TeV}) \lesssim 2.1$ in
the calculation of the inelastic profile function using Eq.~(\ref{eq:profile2}).
Plots of Eq.~(\ref{eq:profile2}) are shown as dotted curves in Figs.~\ref{fig:corr_compare2}(a-b).  In these calculations we 
allow $m_g$ to vary slowly with $x$ and $p_t^c$ in accordance with the parametrization in Ref.~\cite{Frankfurt:2003td}.  
The total inclusive dijet cross section is the same as what is used 
in~\cite{Rogers:2008ua}, based on the CTEQ6M gluon distribution 
function~\cite{Pumplin:2002vw} with a $K$ factor of $1.5$.
With next-to-leading-order PDFs being used, the $K$ factor is closer to $1.2$.
However the dominant contribution to final states typically involves at least 
three jets, corresponding to $K=1.5$ for our calculation~\cite{vogelsang}.  See also the discussion 
in Ref.~\cite{Rogers:2008ua}.

In Fig.~\ref{fig:corr_compare2}(a), we have used $\eta_4 \approx 1.3$ while 
in Fig.~\ref{fig:corr_compare2}(b)we have used $\eta_4 \approx 2.1$.  
In addition, we have tested the sensitivity to higher correlations by including terms up to $n=4$ in Eq.~(\ref{eq:general}), 
and using the approximation $\eta_8 \approx \eta_6 \approx \eta_4 \equiv \eta$.  
The resulting curves are shown as dashed lines in Figs.~\ref{fig:corr_compare2}(a-b).  
The suppression at large $b$ from double correlations could in principle be spoiled if 
the triple correlation is large and positive, so we have checked the case where $\eta_6 > 0$.
Then, to avoid the possibility that $\Gamma > 1$ at very small $b$, $\eta_8$ is made positive.

For comparison, the completely 
uncorrelated eikonal-type expression, Eq.~(\ref{eq:profile}), is plotted as a dash-dotted curve in Figs.~\ref{fig:corr_compare2}(a-b).
Note that there is a substantial difference between the correlated curves and the uncorrelated 
expression for $0.8$~fm$\lesssim b \lesssim 2$~fm in both plots. 
In all cases, the correlations result in a suppression of the total inelastic cross section from dijets [calculated 
by integrating $\Gamma^{\rm inel}_{\rm dijets}(s,b)$ over $b$] by more than $15\%$. 
The shaded regions mark the area covered by the standard extrapolations of the total inelastic profile function.
They correspond to Eq.~(\ref{eq:totparam}) with the range of parameters in Eqs.~(\ref{eq:params1},~\ref{eq:params2}).  
The $p_t$ cutoff in all cases is fixed at the typical value of $p_t^c \approx 2.5$~GeV.   

The uncorrelated curve lies entirely above the shaded area for $b \lesssim 1.6$~fm, in violation of Eq.~(\ref{eq:consistency}).
That is, the hard contribution to the total inelastic cross section is larger than the total inelastic cross section itself for much of the essential 
range of impact parameters.  The curves that include double or quadratic correlations exhibit greater consistency for the 
full range of $b$ for both Figs.~\ref{fig:corr_compare2}(a) and~\ref{fig:corr_compare2}(b).
In the case of the moderate sized correlation corrections in Fig.~\ref{fig:corr_compare2}(a), the effect of 
including triple and higher correlation corrections is rather small compared with the case where only corrections 
from double correlations are kept.  Including higher correlation corrections does seem to smooth out the shape of the profile 
function.  (We have also assumed all correlation corrections to be positive.)  However, the higher correction terms only become significant 
at small impact parameters where the profile function is already close to unity anyway.
If the correlation corrections are larger, as in Fig.~\ref{fig:corr_compare2}(b), then the higher $n > 2$ correlations are more significant. 

\begin{figure*}
\centering
\includegraphics[scale=.45]{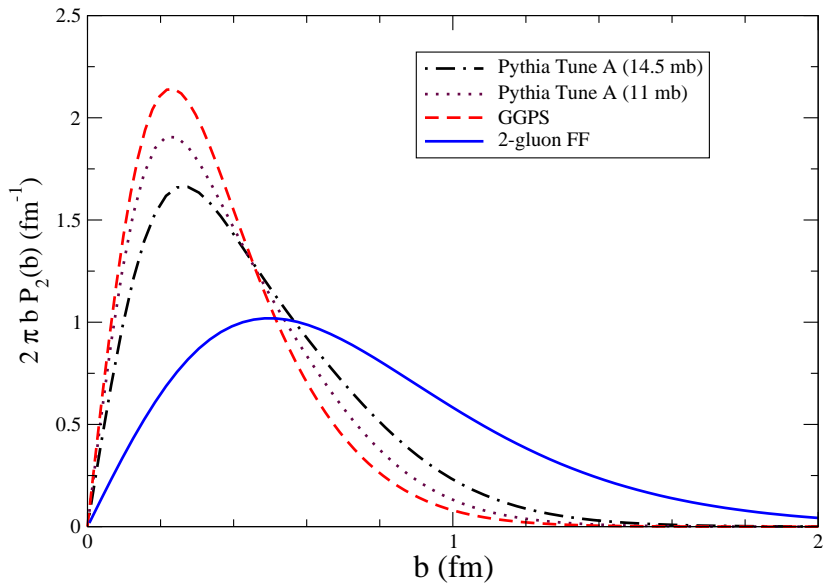}
\caption{Overlap functions obtained from the two gluon form factor,
Eq.~(\ref{eq:FSWoverlap}) (solid curve), the Godbole-Grau-Pancheri-Srivastava (GGPS) model Ref.~\cite{GPS} (dashed curve), and 
the Pythia Tune A overlap function Eq.~(\ref{eq:pythia}) fitted to $\sigma_{\rm  eff} = 14.5$~mb (dash-dotted curve) 
and $\sigma_{\rm eff} = 11$~mb (dotted curve).}
\label{fig:overlaps}
\end{figure*}

Now let us consider what is needed for the radius of the hard overlap function 
if all the $\eta_{2n}(s)$ are set to zero in Eq.~(\ref{eq:general}) 
(reducing to the standard eikonal formula).  If a small value of $p_t^c$ is used 
to evaluate $\sigma_{\rm pQCD}^{inc}(s;p_t^c)$, 
then fits of the total cross section to current data require a very narrow width for the overlap function~\cite{Engel:2001mm}.  
In theoretical calculations, a narrow overlap function is obtained, for example, 
in the semiperturbative approach proposed in~\cite{GPS} where the radius of the hard overlap function decreases with energy.
In Pythia the hard overlap function is modeled by the double Gaussian parametrization~\cite{Sjostrand:2004pf}
\begin{multline}
\label{eq:pythia}
P_2(b) = \frac{(1-\beta)^2}{2 a_1^2} \exp \left\{ \frac{-b^2}{2 a_1^2} \right\} + \\ + \frac{2 \beta (1-\beta)}{a_1^2 + a_2^2} 
\exp \left\{ \frac{-b^2}{a_1^2 + a_2^2} \right\} 
+ \frac{\beta^2}{2 a_2^2} \exp \left\{ \frac{-b^2}{2 a_2^2} \right\},
\end{multline}
with $a_2 = 0.4 \, a_1$ and $\beta = 0.5$ (in Tune A).  The radius in Eq.~(\ref{eq:pythia}) does not vary with energy. 
We determine $a_1$ by using Eq.~(\ref{eq:pythia}) in Eq.~(\ref{eq:sigeffuncor}) 
for $\sigma_{\rm eff}$ with uncorrelated multiple hard scattering and fixing it to the measured CDF value.  
In Fig.~\ref{fig:overlaps} we show Eq.~(\ref{eq:pythia}) with $a_1$ 
calculated using $\sigma_{\rm eff}=14.5$~mb~(Ref.~\cite{Abe:1997xk}) and $\sigma_{\rm eff}=11$~mb~(Ref.~\cite{Treleani:2007gi}).  
For comparison we have also
plotted the overlap function Eq.~(\ref{eq:FSWoverlap}) obtained from the two-gluon form factor at $\sqrt{s} = 14$~TeV, 
and the overlap function from Ref.~\cite{GPS}, also 
at $\sqrt{s} = 14$~TeV~\footnote{R. Godbole and G. Pancheri provided  
data tables for their model of the hard overlap function in this comparison.}.  
A comparison of these curves illustrates that models of 
the overlap function tend to be much narrower than what is expected from the two-gluon form factor. 

Because $a_1$ and $\eta_{2}$ are fixed by the measured value of $\sigma_{\rm eff}$, Eq.~(\ref{eq:pythia}) 
yields by construction the same rate of double 
collisions as Eq.~(\ref{eq:profile2}).  
For triple and quadruple collisions the two approaches give roughly similar rates (same orders of magnitude).   
Therefore, in practical situations, using Eq.~(\ref{eq:pythia}) with a narrow peak may be an economical way to  
model the effects of correlations.
However, our basic aim in this paper is to 
incorporate the maximum amount of available experimental input 
into the description of hard collisions by using the factorization theorem and
parametrizations of the GPD to describe the overlap function.
Using the overlap function obtained directly from the $t$ dependence of the $J/\psi$ photoproduction cross section
requires either that a larger transverse momentum cutoff ($p_t^c \gtrsim 3.5$~GeV) is used, or that
double correlations are incorporated by using Eq.~(\ref{eq:profile2}) with $\eta_{2n} > 0$.
Otherwise, there are potential problems with the consistency relation Eq.~(\ref{eq:consistency}), even for a relatively large $b$.

%%%%%%%%%%%%%%%%%%%%%%%%%%%%%%%%%%%
%%%%%%%%%%%%%%%%%%%%%%%%%%%%%%%%%%%
%%%%%%%%%%%%%%%%%%%%%%%%%%%%%%%%%%%
\begin{figure*}
\centering
\begin{tabular}{c}
    \includegraphics[scale=0.4]{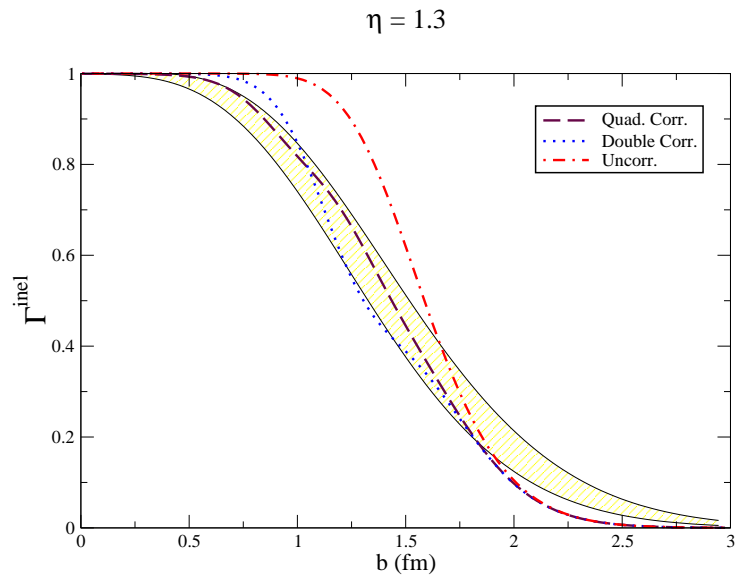}
  \\[3mm]
  (a)
  \\[10mm]
    \includegraphics[scale=0.4]{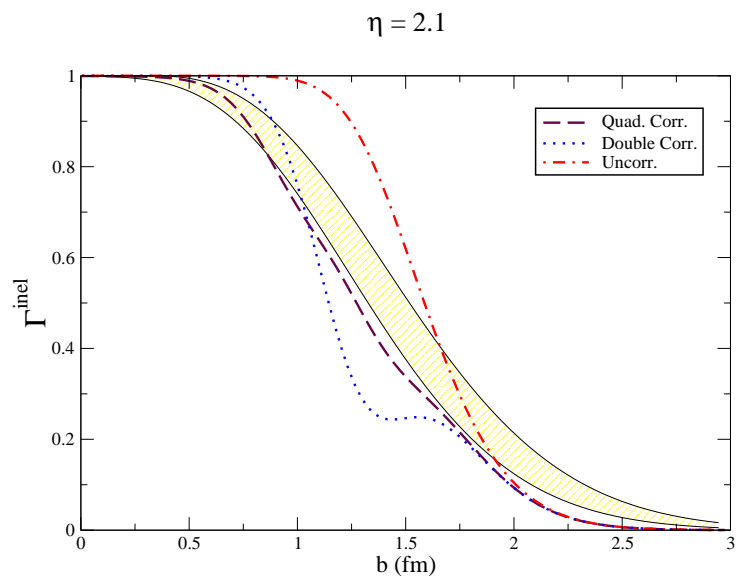}
  \\[3mm]
  (b)
  \end{tabular}
\caption{Inelastic profile functions calculated exactly as in Fig.~\ref{fig:corr_compare2gaus} but now 
with the Gaussian form for the overlap function.}
\label{fig:corr_compare2gaus}
\end{figure*}
%%%%%%%%%%%%%%%%%%%%%%%%%%%%%%%%%%%
%%%%%%%%%%%%%%%%%%%%%%%%%%%%%%%%%%%
%%%%%%%%%%%%%%%%%%%%%%%%%%%%%%%%%%%
One source of uncertainty is the shape of the overlap function $P_2(s,b;p_t^c)$.
A Gaussian form, for instance, may be preferred to Eq.~(\ref{eq:FSWoverlap}).
Therefore, we have repeated the calculation of Fig.~\ref{fig:corr_compare2}, but 
now with 
\begin{equation}
\label{eq:gauss}
P_2(s,b;p_t^c) = \frac{1}{2 \pi b_0} \, \exp \left[ \frac{-b^2}{2 b_0} \right].
\end{equation}
The parameter $b_0$ is fixed by requiring that the average $b^2$,
\begin{equation}
\label{eq:avb}
\langle b^2 \rangle = \int d^2 {\bf b} \, b^2 P_2(s,b;p_t^c),
\end{equation}
is the same for both Eq.~(\ref{eq:gauss}) and Eq.~(\ref{eq:FSWoverlap}).
The resulting plots are shown in Fig.~\ref{fig:corr_compare2gaus}.
The drop with $b$ is slightly steeper at intermediate $b$ in Fig.~\ref{fig:corr_compare2}, 
but otherwise the plots are very similar.
We also point out that a recent experimental study~\cite{Collaboration:2009xp} finds 
good agreement between $\rho$ and $\phi$ electroproduction data and the dipole form for the 
two-gluon form factor.
%%%%%%%%%%%%%%%%%%%%%%%%%%%%%%%%%%%%%%%%%%%%

Another source of uncertainty is the contribution from diffraction to the inelastic 
cross section, which is expected to be much more peripheral than generic inelastic interactions.  
This is known already from analyses of the diffractive 
processes at lower energies~\cite{Alberi:1981af} and should be 
even more prominent at $\sqrt{s} = 2 \, {\rm TeV}$  and above where 
inelastic diffraction cannot occur at small impact parameters, 
and where the interaction is practically black.  
Inelastic diffraction constitutes a significant 
fraction of the inelastic cross section at $\sqrt{s} = 2 \, {\rm TeV}$, 25\% - 30  \%,  
and is expected to remain significant at the LHC. 

Hence, in the region where we use the consistency requirement, 
Eq.~(\ref{eq:consistency}), a large fraction of $\Gamma^{\rm inel}$ is due to inelastic diffraction.
At the same time the Tevatron data on the jet production in diffraction indicate a large 
suppression of jet production as compared to expectations based on the use of the 
diffractive PDFs measured at HERA. This indicates that for  the values of $p_t^c$ which we discuss the fraction of the 
inelastic diffractive events with jets is significantly smaller than $1$.  Hence,  
Eq.~(\ref{eq:consistency}) should be applied to $\Gamma^{\rm inel}(b) - \Gamma^{\rm diff}(b)$.
This implies that an even larger $p_t$ cutoff is needed. A more quantitative analysis of 
this effect requires modeling of the inelastic diffraction profile function and of the 
dynamics of the suppression of jet production in diffraction. We leave such investigations to future work.
%%%%%%%%%%%%%%%%%%%%%%%%%%%%%%%%%%%%%%%%%%%

\section{Tests of Impact Parameter Dependence}
\label{sec:impdep}

\begin{figure*}
\centering
\includegraphics[scale=.4]{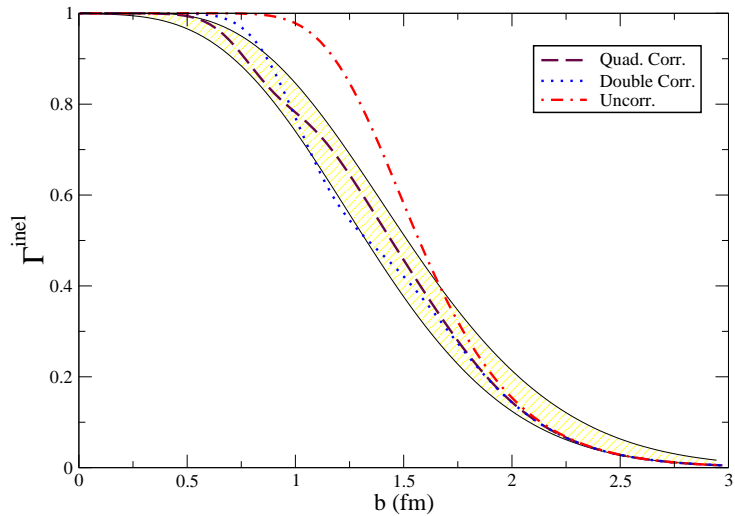}
\caption{Curves analogous to Fig.~\ref{fig:corr_compare2}(b), but now using the $b$-dependent 
correlation correction in Eq.~(\ref{eq:bdep}).}
\label{fig:bdep}
\end{figure*}

A large source of uncertainty is in the role of $b$ dependence in the correlation 
corrections.  
As far as we know, there are currently no predictions of the impact 
parameter dependence of nonperturbative correlations in multiparton distributions.
If all correlations are localized at small impact parameters, then at large impact parameters
one simply recovers the 
uncorrelated model.  We find such scenarios unlikely, however, since 
the binding interaction between any constituent partons should be expected to be large, regardless 
of impact parameter.  
By relaxing the assumption of $b$ independence for 
all correlation corrections, it is possible to 
reconstruct arbitrarily different shapes for the profile function, though in principle 
this arbitrariness can be reduced by future measurements of higher correlations.
In Figs.~\ref{fig:corr_compare2}(b) and~\ref{fig:corr_compare2gaus}(b), the dip 
at intermediate $b$ that occurs when only double correlations are included suggests
that higher correlations should be included.  However, a smooth form for the inelastic 
profile function can also be recovered if we allow for modest impact parameter dependence
in the double correlation correction.  As an example, we use instead of $\eta_4 = 2.1$, 
\begin{equation}
\label{eq:bdep}
\eta_4(b) = e^{-2 b^2} + \frac{3.5}{2 + b}.
\end{equation}
This gives numerically the same $\sigma_{eff}$ as Eq.~(\ref{eq:sigeffcor}) with $\eta_4 = 2.1$, 
but now with correlations peaked at small $b$ and with a weakly falling tail at large $b$.
The plot analogous to Fig.~\ref{fig:corr_compare2}(b) is shown in Fig.~\ref{fig:bdep}.
The curve with only double correlations is now closer to the corresponding curve
in Fig.~\ref{fig:corr_compare2}(b) which was for a larger $\sigma_{eff}$.
However, the main conclusion of the previous section remains valid --- that correlation 
corrections of size roughly $1$ to $2$ are needed even at relatively large $b$ in order to have 
consistency with Eq.~(\ref{eq:consistency}).
%%%%%%%%%%%%%%%%%%%%%%%%%%%%%%%%%%%%%%%%%%%%%%%%%%%%%%%%%%%%
%%%%%%%%%%%%%%%%%%%%%%%%%%%%%%%%%%%%%%%%%%%%%%%%%%%%%%%%%%%%

It is possible to visualize why an enhancement in correlations at large impact parameters
is likely 
by considering the following simple model:  Consider scattering at a large impact 
parameter $|{\bf b}| = \rho$ and assume that  
collisions are between the pion clouds in each nucleon with the core of the other nucleon. 
Then the cross section for scattering in a range of impact parameters from $\rho - r_\pi$ to $\rho + r_\pi$ 
(where $r_\pi$ is the pion radius) is proportional to the 
cross-sectional area of the pion and the probability ${\rm prob}_\pi(\rho)$ to scatter if the pion cloud from one 
nucleon overlaps with the nucleon core of the other:
\begin{equation}
d \sigma_{\rm cor} \propto \pi r_\pi^2 \times {\rm prob}_\pi(\rho). 
\end{equation}
By contrast, without correlations the cross section is 
proportional to the area of the annulus between $\rho - r_\pi$ and $\rho + r_\pi$
and a different probability,
\begin{equation}
d \sigma_{\rm uncorr} \propto  4 \pi \rho r_\pi \times {\rm prob}_{\rm uncorr}(\rho). 
\end{equation}
Requiring that these two expressions give the same cross section means that
\begin{equation}
\frac{{\rm prob}_\pi(\rho)}{{\rm prob}_{\rm uncorr}(\rho)} = \frac{4 \rho}{r_\pi}.
\end{equation}
The cross sections for double scattering are the areas times of the 
square of the probabilities for single scattering.  The ratio of the double scattering 
cross sections obtained for the correlated and uncorrelated cases is then
\begin{equation}
\frac{\pi r_\pi^2 [ {\rm prob}_\pi(\rho)]^2}{4 \rho \pi r_\pi [{\rm prob}_{\rm uncorr}(\rho)]^2} = \frac{4 \rho}{r_\pi}.
\end{equation}
Hence, in this simple picture one expects the true double scattering cross section to be enhanced at large $\rho$
relative to what is expected if correlations are neglected.
%%%%%%%%%%%%%%%%%%%%%%%%%%%%%%%%%%%%%%%%%%%%%%%%%%%%%%%%%%%%
%%%%%%%%%%%%%%%%%%%%%%%%%%%%%%%%%%%%%%%%%%%%%%%%%%%%%%%%%%%%

A natural question is how to test the dependence of correlations between 
partons as a function of $b$. One possibility is to study the dependence of 
multijet production on the associated hadron multiplicity away from the 
rapidities and angles where hadron production due to the fragmentation of jets is important.
The distribution in $b$ of events with dijets is given by the overlap function $P_2(b)$ [Eq.~\ref{eq:prob_one}]. 
Let us now consider the distribution over the accompanying multiplicity $P_M(N/ \langle N \rangle)$ 
where $N$ is the observed hadron multiplicity, and $\langle N \rangle$ is the average multiplicity in 
minimum bias nondiffractive events (the CDF collaboration reported an average 
multiplicity for events with a $Z$ boson trigger for angles where gluon radiation 
effects associated with the production of $Z$ are small by a factor $\approx 2$ larger than 
in the minimal bias inelastic events~\cite{Acosta:2004wqa}).
%%%%%%%%%%%%%%%%%%%%%%%%%%%%%%%%%%%
%%%%%%%%%%%%%%%%%%%%%%%%%%%%%%%%%%%

Both soft and hard interaction mechanisms of hadron production lead to a 
monotonic increase in the average accompanying multiplicity with decreasing $b$. 
In the first case this is due to an increase of the contribution of the multi-Pomeron 
exchanges; in the second it is due to an increase in the probability of multiple 
(semi)hard interactions.  Hence, to a large extent fluctuations in the multiplicity are
due to the distribution of collisions in $b$. To simplify the following 
discussion we will neglect fluctuations of multiplicity at a given $b$, 
though one can model these effects as well
using current Monte Carlo  models. It seems natural to expect that such fluctuations will smear 
resolution in $b$ but will not remove the generic trend 
of decrease of the multiplicity with increase of $b$.
 
Under these assumptions,  
we can identify intervals of $b$ that correspond to production of 
$i$ hadrons not belonging to the hard collision (this could be, for example, hadrons at the central rapidity interval not affected by hadron production in the hard process leading to production of dijets), 
  \begin{equation}
  \int_{b_{i+1}}^{b_i} P_2(b)d^2b=p_i.
  \label{P2Mdi}
  \end{equation}
where $p_i$ is the probability for producing exactly $i$ hadrons.
Given values of $p_i$ and $P_2(b)$, we may construct the number of hadrons $N(b)$ corresponding to a given impact parameter:
\begin{eqnarray}
N(b) & = 1 \qquad {\rm for} \, b \in \left[ b_2, b_1 \right], \\
N(b) & = 2 \qquad {\rm for} \, b \in \left[ b_3, b_2 \right], \\
\vdots \nonumber
\end{eqnarray}
   
Next we can consider production of four jets.  We may calculate the multiplicity of events with 
at least four jets by integrating the number of collisions $N(b)$ over $b$, weighted by the probability density $P_4(b)$ for 
a four-jet event:
\begin{equation}
\langle N^{(4)} \rangle = \int_0^{\infty } P_4(b) N(b) d^2b. 
\label{P4M}
\end{equation}     
One can write similar equations for the higher moments of $N^{(4)}$, 
though in this case sensitivity to the fluctuations at a given impact parameter becomes larger.

If correlation corrections are impact parameter independent, then $P_4(b)$ is simply proportional to the 
square of $P_2(b)$.  However, if correlations are concentrated at small impact parameters, then 
$P_4(b)$ is more sharply peaked at small $b$ and falls off faster at large $b$.  
Then the integrand in Eq.~(\ref{P4M}) is more localized at small $b$ where $N(b)$ is large, and 
the average accompanying multiplicity for four-jet events will be larger than in the uncorrelated case. 
A similar analysis extends to higher moments $\langle N^{(i)} \rangle$.
One should emphasize here that the multiplicity of hadron production in the rapidity 
interval between $4$ jets may be affected by various effects of color correlations, etc.;
hence it is  desirable to look for the change of the multiplicity at the rapidities 
sufficiently  remote from the region of 4 -jet activity. This is feasible for the 
LHC detectors with a good acceptance in a large rapidity interval.

In addition, if the correlations are present at all impact parameters they should be 
manifested in the hard diffractive processes which correspond to scattering 
at large impact parameters.  One could consider both cases of 
single and double diffraction with production of two and four jets:
\begin{eqnarray}
 pp & \to p + X (2 jets +Y, 4 jets +Y);  \\
 pp & \to p p + X (2 jets +Y, 4 jets +Y).
\label{diff}
 \end{eqnarray} 
 
\begin{figure}
\centering
\includegraphics[scale=.37]{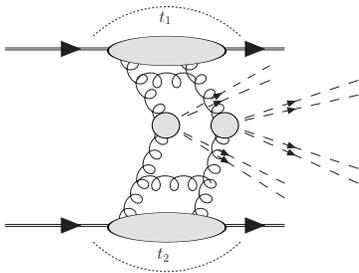}
\caption{Double Pomeron process with production of two pairs of dijets.}
\label{fig:diffraction}
\end{figure}

Correlations between the partons should also enhance the cross section of the exclusive 
channel when the light-cone fraction carried by two of the interacting partons of one of the 
nucleons is close to maximal: $ (x_1+x_2) / x_{\Pomeron} \sim 1$. Such a contribution 
should be enhanced if $-t$ is large enough (few GeV$^2$) to squeeze 
the transverse size of the exchanged ladder (see Fig.~\ref{fig:diffraction}).
%%%%%%%%%%%%%%%%%%%%%%%%%%%%%%%%%%% 
%%%%%%%%%%%%%%%%%%%%%%%%%%%%%%%%%%%

\section{Discussion and Conclusions}
\label{sec:con}

The main conclusion of this paper is that, given the distribution of hard partons known from the gluon GPD, 
the hard contribution to the total inelastic profile function should probably be modeled using Eq.~(\ref{eq:profile2}) or 
Eq.~(\ref{eq:general}) with $1 \lesssim \eta \lesssim 2$ rather than the usual one-minus-exponential shape that arises in a purely eikonal treatment.
This avoids conflicts with general expectations for 
the total inelastic profile function (essentially a unitarity problem), when a small or fixed $p_t^c$ is used for 
the inclusive dijet cross section.  The correlation corrections stem from a breakdown in the factorization ansatz of Eq.~(\ref{eq:badapprox}).

There are many uncertainties associated with the size of correlations at various impact parameters.
Nevertheless, we believe there is ample evidence that large correlations are important even at relatively large impact parameters.
Our sample calculations illustrate how including correlation corrections of roughly the size suggested by available experimental
data can lead to greater consistency with the total inelastic $pp$ cross section.
Hence, questions of unitarization and consistency at small $p_t^c$ should be organized around a more precise
determination of higher correlation corrections, with either experimental input or theoretical modeling.

As more data become available, and the range of allowed parameters narrows, it will be possible to 
use Eq.~(\ref{eq:general}) to
obtain an increasingly refined picture of 
the profile function.  More studies of the $x$ dependence and rate of growth of the hard overlap function are needed.
Furthermore, it will be important to establish, through measurements or theoretical models, the $p_t^c$ and energy dependence of 
the correlation corrections~\cite{Domdey:2009bg}.  
Although we have assumed that correlation corrections are impact parameter independent in this paper, 
it is possible to incorporate any impact parameter dependence into the same basic framework simply by 
allowing the $\eta_{2n}(s)$ to depend on impact parameter.  
We have suggested 
possible ways of testing for impact parameter dependence of correlations in Sec.~\ref{sec:impdep}.
It would also be very useful to have direct measurements of contributions from triple and higher correlations.  
For a recent discussion of this possibility 
for triple correlations at the LHC, see Ref.~\cite{Maina:2009vx}.
Furthermore, it was recently argued in Ref.~\cite{Calucci:2009ea} that different numbers of 
collisions contribute incoherently when they are identified by their topologies.

Fits to the total cross section can be obtained within the more common eikonal approach by using a very narrow hard overlap function.  
However, the contribution to the total cross section arising from high $p_t$ jets depends on the blackness of the 
hard collisions~\cite{Frankfurt:2003td,Frankfurt:2006jp}, and hence on the width of the hard overlap function.  
Therefore, it is phenomenologically important 
to use the correct radius in the description of hard multiple collisions. 
In addition to being necessary for the construction of realistic simulations, sorting out 
these issues has the potential to 
lead to an improved understanding of the transverse structure of the proton at high energies.

\section*{Acknowledgments}

We thank P. Skands for discussions of the Pythia model of the overlap function.
We thank A. Sta\'sto and D. Treleani for useful discussions.  We also thank R. Godbole and G. Pancheri 
for providing us with data for their overlap function.
Figures were made using Jaxodraw~\cite{jaxo}.
T.C. Rogers was supported by the research program of the ``Stichting voor Fundamenteel Onderzoek der Materie (FOM)'', 
which is financially supported by the ``Nederlandse Organisatie voor Wetenschappelijk Onderzoek (NWO)''.
M. Strikman was supported by the U.S. Department of Energy under Grant No. DE-FG02-93ER-40771.

\end{document}